\documentclass[12pt,tightenlines,nofootinbib,superscriptaddress, APS]{revtex4}

\usepackage{amsfonts,amssymb,amsthm}
\usepackage{appendix}

\usepackage{amssymb,amstext,amsmath,amsthm}
\usepackage{amsfonts}
\usepackage{color}
\usepackage{float}
\usepackage{bbold}
\usepackage{indentfirst}
\usepackage{tabu}
\usepackage{verbatim}
\usepackage{comment}

\newcommand{\be}{\begin{equation}}
\newcommand{\ee}{\end{equation}}
\newcommand{\barray}{\begin{array}}
\newcommand{\earray}{\end{array}}
\newcommand{\bea}{\begin{eqnarray}}
\newcommand{\eea}{\end{eqnarray}}
\newcommand{\bs}{\begin{subequations}}
\newcommand{\es}{\end{subequations}}
\newcommand{\balign}{\begin{align}}
\newcommand{\ealign}{\end{align}}
\newcommand{\equ}{\begin{equation}}
\newcommand{\nequ}{\end{equation}}
\newcommand{\eqa}{\begin{eqnarray}}
\newcommand{\neqa}{\end{eqnarray}}

\def\sig{\sigma}
\def\ph{\varphi}

\def\bs{\bar{\sigma}}

\def \mp {{\mu'}}

\newcommand{\om}{\ominus}
\newcommand{\op}{\oplus}

\newcommand{\rd}{\mathrm{d}}

\def\sl2c{SL(2,\mathbb{C})}

\newcommand{\mc}[1]{\mathcal{#1}}

\def\bs{\bar{s}}

\usepackage{bbm}

\usepackage{tikz}
\usetikzlibrary{decorations.pathmorphing}
\usepackage{tkz-euclide}
\usetikzlibrary{decorations.markings}
\tikzset{->-/.style={decoration={
  markings,
  mark=at position .5 with {\arrow{>}}},postaction={decorate}}}

\begin{document}

\title{\Large{Scalar Field Theory in Curved Momentum Space}}

\author{{Laurent Freidel}}
\smallskip 
\affiliation{{Perimeter Institute for Theoretical Physics, Waterloo, Ontario, Canada}}
\author{{Trevor Rempel}}

\affiliation{{Perimeter Institute for Theoretical Physics, Waterloo, Ontario, Canada}}
\affiliation{\small\textit{Department of Physics, University of Waterloo, Waterloo, Ontario, Canada}}
\smallskip
\date{ \today}


\bigskip
\begin{abstract}
We derive an action for scalar quantum field theory with cubic interaction in the context of relative locality. Beginning with the generating functional for standard $\ph^3$--theory and the corresponding Feynman rules we modify them to account for the non--trivial geometry of momentum space. These modified rules are then used to reconstruct the generating functional and extract the action for the theory. A method for performing a covariant Fourier transform is then developed and applied to the action. We find that the transformed fields depend implicitly on a fixed point in momentum space with fields based at different points being related by a non-local transformation. The interaction term in the action is also non--local, but the kinetic term can be made local by choosing the base point to be the origin of momentum space.
\end{abstract}
\maketitle

\section{Introduction}
Relative locality (see \cite{relativelocality},\cite{gammarayburst},\cite{deepening}) considers a paradigm in which one systematically weakens the notion of absolute locality by allowing momentum space to posses a non--trivial geometry. In this framework momentum space is taken to be fundamental and spacetime emergent from the geometric structure of momentum space. This geometry is codified in terms of a metric and a connection which measure modifications in the energy--momentum relations and non--linearities in the conservation law, respectively. The most startling prediction of the theory is that localization of events becomes an observer dependent phenomenon, with the degree of the non--locality scaling with an observers distance from the event.\\
\indent As originally formulated relative locality describes a ``classical non--gravitational'' regime in which $\hbar$ and $G_\mathrm{N}$ are neglected but their ratio $m_p=\sqrt{\hbar/G_\mathrm{N}}$ is held fixed. In neglecting $\hbar$ the theory offers no formulation of quantum field theory and therefore no insight into particle phenomenology. The goal of this paper is to take a first step towards addressing this issue by ``turning $\hbar$ back on.'' More specifically, we derive an action for scalar field theory with cubic interaction term in the framework of relative locality. \\
\indent We begin with the generating functional for standard $\ph^3$--theory, Fourier transform this into momentum space and extract the corresponding Feynman rules. We then deform these rules to account for the non--trivial geometry on momentum space. With modified momentum space Feynman rules in hand we write down the corresponding generating functional and read off the action for our theory. The action will be written in terms of momenta and should be Fourier transformed into spacetime. However, since momentum space is curved any transformation we perform should preserve the covariance of the action. As such, it will be necessary to develop a method for performing a covariant Fourier transform. We develop this method in detail and then apply it to our action. In doing so we find that the transformed fields depend, implicitly, on a fixed point in momentum space with fields based at different points being related by a non--local transformation. This implies that there are a continuum of quantum field theories, one for each point in momentum space. The transformed action is also non-local, although the kinetic term can be made local by choosing the base point to be the origin of momentum space. In this case the relative locality action is of the form 
\begin{align}
S_{RL} &=\frac{1}{2}\int d\nu(x) \left[\left(\hat{\ph} \Box \hat{\ph}\right)(x) - m^2 \hat{\ph} \hat{\ph} (x)\right] + \frac{g}{3!}\int d\nu(x) \left(\hat{\ph} \star \left(\hat{\ph} \star \hat{\ph}\right)\right)(x),
\end{align}
where $\star$ denotes a non commutative and non associative product that encodes the non trivial geometry of momentum space via its deformed addition.

\section{Geometry of Momentum Space}
In what follows we take momentum space to be a non--linear manifold $\mathcal{P}$ and phase space the cotangent bundle $T^*\mathcal{P}$. Spacetime then emerges as cotangent planes to points in momentum space $T^*_p\mathcal{P}$. We will now embark on a self--contained review of momentum space geometry; the presentation will be as general as possible, although in later sections we will be forced to give up some of this generality for the sake of coherence and ease of calculation.

\subsection{Combination of Momenta}
Any process in which particles interact with each other is governed by imposing conservation of energy and momentum.
Our choice of conservation law is an expression of the locality properties of the theory.
Here we want to investigate conservation laws which are non linear and understand the effects such a modification has on the formulation of field theory.
Thus, we postulate a rule, $\op$, for combining particles momenta. Before we define this rule let us pause and consider what properties it should posses. First, interaction with a zero momentum object will produce no change in momenta and so $0$ should be an identity for $\op$. Secondly, we need a method for turning an incoming particle into an outgoing one and so our rule needs an inverse. As mentioned we will not assume this rule is linear and so there is no reason to demand either commutativity or associativity. In keeping the rule as general as possible we allow the physics to tell us what properties are mathematically acceptable. Formally, we define our rule as a $C^{\infty}$ map:
\begin{equation}
\begin{aligned}
\op&: \mathcal{P}\times\mathcal{P} \to \mathcal{P}\\
&\quad (p,q) \mapsto p \op q,
\end{aligned}
\end{equation}
having identity $0$
\begin{align}
0\op p = p \op 0 = p \qquad \forall p \in \mathcal{P},
\end{align}
and inverse $\om$
\begin{align}
(\om p) \op p = p \op (\om p) = 0 \qquad \forall p \in \mathcal{P}.
\end{align} 
Note that we assume a unique inverse; if $p,q \in \mathcal{P}$ are such that $q \op p = p \op q = 0$ then $q = \om p$.  \\
\indent Equipped with this combination rule we can enforce the conservation of energy and momentum at each interaction. We will write this as\footnote{In special relativity  $\mathcal{K}_\mu(p^I) = \sum_I p^I_\mu$}
\begin{align}\label{conserve1}
\mathcal{K}_\mu(p^I) = 0,
\end{align}  
where $I = 1,2,\ldots$ runs over the number of particles participating in the interaction. For example, a process with two incoming particles $p,q$ and one outgoing particle $k$ may have
\begin{align}\label{conserve}
\mathcal{K}_\mu = (p \op (q \om k))_\mu,
\end{align} 
where we have made use of the obvious notation $q \om k = q \op (\om k)$ and have adopted the convention that all momenta are taken to be incoming. Observe that \eqref{conserve} is just one of twelve possible choices for $\mathcal{K}$ all of which are distinct if $\op$ is neither commutative nor associative. Differences arising from alternate choices of the conservation law are explored in detail in \cite{Ricardo}.\\
\indent Suppose we are given a generic conservation law $p \op (q\op k) = 0$. For this to be meaningful it must be possible to solve for any one of the momenta uniquely in terms of the other two. To address this issue we introduce left ($L_p$) and right ($R_p$) translation operators
\begin{align}
L_p(q) \equiv p \op q \qquad \mathrm{and} \qquad R_p(q)\equiv q \op p,
\end{align}
which allow the conservation law to be re--written as
\begin{align}
R_{q\op k}(p) = L_p(R_k(q)) =L_p(L_q(k)) = 0.
\end{align}  
The existence of a unique solution for each momenta then reduces to the requirement that the left and right translation operators be invertible. It is therefore assumed that $L_p^{-1}$ and $R_p^{-1}$ exist for all $p \in \mathcal{P}$ and so the solutions of our conservation law are given by
\begin{align}
p = \om(q\op k)\qquad q = R_k^{-1}\left(\om p\right) \qquad
k = L_q^{-1}\left(\om p\right),
\end{align}
where we have used that $L_p^{-1}(0) = R_{p}^{-1}(0) = \om p$, by the uniqueness of the inverse. Note that we are not assuming the composition law $\op$ is left or right invertible; doing so would be equivalent to setting $L^{-1}_p = L_{\om p}$ and $R^{-1}_p = R_{\om p}$ respectively.

\subsection{Curvature and Torsion}
The algebra induced on momentum space by our composition rule determines a connection on $\mathcal{P}$ via
\begin{align}
\Gamma_\rho^{\mu\nu}(0) = \frac{\partial}{\partial p_\mu}\frac{\partial}{\partial q_\nu} (p \op q)_\rho \Big\vert_{p,q = 0}.
\end{align}
The torsion is the anti-symmetric part of $\Gamma^{\mu\nu}_\rho$ and measures the extent to which the combination rule fails to commute
\begin{align}
T^{\mu\nu}_\rho(0) = \Gamma^{[\mu\nu]}_\rho(0) = \frac{\partial}{\partial p_\mu}\frac{\partial}{\partial q_\nu} (p \op q - q \op p)_\rho \Big\vert_{p,q = 0}.
\end{align}
Similarly, the curvature of $\mathcal{P}$ is a measure of the lack of associativity of the combination rule
\begin{align}
R^{\beta\gamma\delta}_{\phantom{\beta\gamma\delta}\mu}(0) = -2\frac{\partial}{\partial p_{[\beta}}\frac{\partial}{\partial q_{\gamma]}}\frac{\partial}{\partial k_\delta}\left(p \op (q\op k) - (p \op q) \op k\right)_\mu \Big\vert_{p = q = k = 0}. 
\end{align}
Unlike general relativity the connection $\Gamma^{\mu\nu}_\rho$ is not necessarily metric compatible and so $g^{\mu\nu}$ may fail to be covariantly constant. To measure the extent to which the covariant derivative of $g^{\mu\nu}$ deviates from zero we introduce the non-metricity tensor
\begin{align}
N^{\mu\nu\rho} = \nabla^\mu g^{\nu\rho} = \partial^\mu g^{\nu\rho} - \Gamma ^{\nu\mu}_\alpha g^{\alpha \rho} - \Gamma^{\rho \mu}_\alpha g^{\nu\alpha}.
\end{align}
\indent Let $\left\lbrace \begin{smallmatrix} \mu \;\nu \\ \rho \end{smallmatrix}\right\rbrace$ denote the standard Levi-Civita connection compatible with the metric $g^{\mu\nu}$. We can then decompose the full connection $\Gamma^{\mu\nu}_\rho$ in--terms of the Levi--Civita connection, the torsion and the non-metricity tensor, viz
\begin{align}
\Gamma^{\mu\nu}_\rho = \left\lbrace \begin{smallmatrix} \mu \;\nu \\ \rho \end{smallmatrix}\right\rbrace +\frac{1}{2}T^{\mu\nu}_\rho - \frac{1}{2}g_{\rho\alpha}\left(N^{\mu\nu\alpha}+N^{\nu\mu\alpha} - N^{\alpha\mu\nu} + T^{\alpha\mu\nu} + T^{\alpha\nu\mu}\right),
\end{align}
where $T^{\mu\nu\rho} = T^{\mu\nu}_\alpha g^{\alpha\rho}$. Similarly, we can expand the non--metricity tensor in--terms of the torsion and the symmetric tensor  $\mathcal{N}^{\mu\nu}_\rho = \Gamma^{(\mu\nu)}_\rho - \left\lbrace \begin{smallmatrix} \mu \;\nu \\ \rho\end{smallmatrix}\right\rbrace$, the result is
\begin{align}\label{nonmetricity}
N^{\mu\nu\rho} = \frac{1}{2}\left(T^{\mu\nu\rho} + T^{\mu\rho\nu}\right) - \mathcal{N}^{\nu\mu}_\alpha g^{\alpha \rho} - \mathcal{N}^{\rho\mu}_\alpha g^{\alpha \nu}.
\end{align}

\subsection{Transport Operators}
In order to write the locality equations at each vertex we need to introduce transport operators that arise from the infinitesimal transformation of the addition law. 
We define the left transport operator as
\begin{align}
\left(U^q_{p\op q}\right)^\mu_\nu = \left(\rd_q L_p\right)^\mu_\nu = \frac{\partial (p \op q)_\nu}{\partial q_\mu},
\end{align}
and the right transport operator as 
\begin{align}
\left(V^q_{q\op p}\right)^\mu_\nu = \left(\rd_q R_p\right)^\mu_\nu = \frac{\partial (q \op p)_\nu}{\partial q_\mu}.
\end{align}
Here the notation $\rd_{p}f\equiv (\partial_{p_{\mu}}f(p)) d x^{\mu} $ denotes the differential at $p$ of the function $f$.
The most general form of the transport operators, $U^q_k$ and $V^q_k$, from point $q$ to $k$, can be obtained from the ones defined above by setting $p = R_q^{-1}(k)$ and $p = L_q^{-1}(k)$ respectively. It will also be useful to give a name to the derivative of the inverse:
\begin{align}
\left(I^p\right)^\mu_\nu = \left(d_p \om\right)^\mu_\nu = \frac{\partial (\om p)_\nu}{\partial p_\mu}.
\end{align}
It turns out that these operators are not independent and can be related by 
\begin{align}\label{IUV}
V^p_0 = -U^{\om p}_0 I^p.
\end{align}
The proof of this formula is straight forward and requires only the existence of the inverse $\om p$:
\begin{align*}
0 &= \frac{\partial}{\partial p} (p \op (\om p))\\
&= \frac{\partial}{\partial k}(k \op (\om p))\Big\vert_{k = p} + \frac{\partial}{\partial k}(p \op k)\Big\vert_{k = \om p} \frac{\partial \om p}{\partial p}\\
&= V^p_0 + U^{\om p}_0 I^p.
\end{align*}
By considering equations of the form $L_p(L_p^{-1}(q)) = q$ and $R_p(R^{-1}_p(q)) = q$ we can also derive formulas for the derivatives of $L^{-1}_p$ and $R^{-1}_p$:
\begin{align}
\frac{\partial L_p^{-1}(q)}{\partial q} = \left(U^{L_p^{-1}(q)}_q\right)^{-1}, \qquad  \frac{\partial L_p^{-1}(q)}{\partial p} = - \left(U^{L_p^{-1}(q)}_q\right)^{-1}V^p_q,
\end{align}
and
\begin{align}
\frac{\partial R_p^{-1}(q)}{\partial q} = \left(V^{R_p^{-1}(q)}_q\right)^{-1}, \qquad \frac{\partial R^{-1}_p(q)}{\partial p} = -\left(V^{R_p^{-1}(q)}_q\right)^{-1}U^p_q.
\end{align}
\indent Without demanding certain properties of the composition rule we can not say anything further. For the sake of completeness we now now present a collection of results that are applicable if the following conditions on $\op$ are fulfilled:
\begin{itemize}
\item Composition rule is left invertible, i.e. $L_{p}^{-1}=L_{\ominus p}$:
\begin{align*}
\left(U_{p\op q}^q\right)^{-1} = U_q^{p\op q} \qquad \mathrm{and} \qquad V^{\om p}_{\om p \op q}I^p = - U^q_{\om p \op q}V^p_q
\end{align*}
\item Composition rule is right invertible, i.e. $R_{p}^{-1}=R_{\ominus p}$:
\begin{align*}
\left(V_{q \op p}^q\right)^{-1} = V_q^{q \op p} \qquad \mathrm{and} \qquad U^{\om p}_{q\om p}I^p = - V^q_{q\om p }U^p_q
\end{align*}
\end{itemize}

\subsection{Metric and Distance Function}
It is assumed that the metric on momentum space, $g^{\mu\nu}(p)$, is known. It is then a standard result that the distance between two points $p_0,p_1 \in \mathcal{P}$ along a path $\gamma(\tau)$ is given by: 
\begin{align*}
D_\gamma(p_0,p_1) = \int_{a}^{b}\sqrt{g^{\mu\nu}\left(\gamma(\tau)\right)\frac{d\gamma_\mu}{d\tau} \frac{d\gamma_\nu}{d\tau}}d\tau,
\end{align*}
where $\gamma(a)=p_0$ and $\gamma(b) = p_1$. Of all the paths connecting $p_0$ and $p_1$ \textit{geodesics} will be of principle importance, but here we run into trouble. In relative locality, where the non-metricity tensor does not necessarily vanish, there is more than one viable definition of a geodesic, so it is not immediately clear what one means by a ``geodesic.'' This ambiguity is discussed in Appendix \ref{ap:geodesics}, where we argue that the most appropriate definition of a geodesic is a path which extremizes $D_\gamma(p_0,p_1)$. We will adopt this convention for the remainder of the paper and note that if $\gamma$ is a geodesic we write $D_\gamma(p_0,p_1) = D(p_0,p_1)$. \\
\indent The standard definition of a particles mass is by means of the dispersion relation $p^2 = -m^2$. To account for the geometry of momentum space we deform this relation and assume that the mass of a particle with momentum $p$ is related to the geodesic distance from $p$ to the origin, i.e.
\begin{align}
D^2(p) = -m^2,
\end{align}
where we have used the simplified notation $D(p,0) = D(p)$.

\section{$\ph^3$ Scalar Field}
Having completed our review of the geometric structure of momentum space we will now examine how this new paradigm alters our understanding of quantum field theory. In particular we will consider a quantum scalar field theory with cubic interaction term. 
 
\subsection{Modified Feynman Rules}
The starting point for our analysis will be the generating functional, $Z(J)$, for standard $\ph^3$--theory:
\begin{align}\label{Zpos}
Z(J) = \int\mathcal{D}\ph \exp\left(i\int d^4x \left[-\frac{1}{2}\partial^\mu \ph \partial_\mu \ph - \frac{1}{2} m^2 \ph^2 + \frac{1}{3!}g\ph^3 + J\ph\right]\right).
\end{align}
This is the position space representation of $Z(J)$ which is ill-suited for our purposes. Relative locality treats momentum space as fundamental and so we should Fourier transform $Z(J)$ so that all integrals are over momenta. Denote by $\mathcal{F}$ the Fourier transform of the argument of the exponential, then\footnote{Normally we would denote the Fourier transformed fields as $\hat{\ph}(p)$, $\hat{J}(p)$ but since we will be regarding the momentum space representation as fundamental we will drop the hat.}
\begin{align*}
\mathcal{F} &= i\int \frac{d^4p}{(2\pi)^4} \left(-\frac{1}{2}\left(p^2 + m^2\right)\ph(p)\ph(-p) + J(p)\ph(-p)\right)\\
&\qquad + i\frac{(2\pi)^4g}{3!}\int \frac{d^4p}{(2\pi)^4}\int \frac{d^4q}{(2\pi)^4} \int \frac{d^4k}{(2\pi)^4}\delta(p+k+q)\ph(p)\ph(q)\ph(k)\\
\end{align*}
Following the standard procedure we extract the interaction terms from $Z(J)$ and re-write them as functional derivatives with respect to $J$ acting on the remainder of $Z(J)$. We can then separate out the $J$ dependent terms from the functional by completing the square, in the end we find
\begin{equation}\label{Zmomeq}
\begin{aligned}
Z(J) &= \exp\left(-\frac{(2\pi)^4g}{3!}\int d^4p\int d^4q\int d^4k \delta(p+q+k)\frac{\delta}{\delta J(p)}\frac{\delta}{\delta J(q)}\frac{\delta}{\delta J(k)}\right)\\
&\quad \times \exp\left(\frac{i}{2}\int \frac{d^4p}{(2\pi)^4}J(p)\left(p^2 + m^2\right)^{-1}J(-p)\right)\\
&\qquad \times \int\mathcal{D}\ph \exp\left(-\frac{i}{2}\int \frac{d^4p}{(2\pi)^4}\left(p^2+m^2\right)\ph(p)\ph(-p)\right).
\end{aligned}
\end{equation}
\indent Having successfully removed all $J$ dependence from the functional integral we can evaluate it to obtain some C--number. However, if we insist on the normalization $Z(0) = 1$ we can ignore this number and simply impose the normalization by hand. Hence, 
\begin{equation}\label{Zmom}
\begin{aligned}
Z(J) &\propto \exp\left(-\frac{(2\pi)^4g}{3!}\int d^4p\int d^4q\int d^4k \delta(p+q+k)\frac{\delta}{\delta J(p)}\frac{\delta}{\delta J(q)}\frac{\delta}{\delta J(k)}\right)\\
&\qquad \times \exp\left(\frac{i}{2}\int \frac{d^4p}{(2\pi)^4}J(p)\left(p^2 + m^2\right)^{-1}J(-p)\right).
\end{aligned}
\end{equation}
This generating functional can now be expanded as a sum of all possible Feynman diagrams having $E$ external points, $P$ propagators and $V$ vertices where $E =  3V- 2P$. Each diagram is then assigned a value by means of the following Feynman rules:
\begin{enumerate}
\item To each propagator, \qquad  \begin{tikzpicture}[baseline={([yshift={-3.5ex}]current bounding box.north)}]
\draw[->-, very thick] (0,0) to (2,0);
\draw (1,0) node[anchor=south] {$p$};
\end{tikzpicture}
\quad $=\dfrac{i}{(2\pi)^4(p^2 + m^2)}$;
\item To each external point, 
\qquad  \begin{tikzpicture}[baseline={([yshift={-3.5ex}]current bounding box.north)}]
\draw[->-, very thick] (2,0) to (0,0);
\draw (1,0) node[anchor=south] {$p$};
\draw (0,0) node[circle, draw= black, fill=black, inner sep =2.5pt] {};
\end{tikzpicture}
\quad $=J(p)$;
\item To each vertex, \qquad  \begin{tikzpicture}[baseline={([yshift={-7ex}]current bounding box.north)}] 
\draw[->-, very thick] (-0.707,-0.707) to (0,0);
\draw[->-, very thick] (0.707,-0.707) to (0,0);
\draw[->-, very thick] (0,1) to (0,0);
\draw (-0.707,-0.707) node[anchor=south] {$q$};
\draw (0.707,-0.707) node[anchor=south] {$k$};
\draw (0,1) node[anchor=east] {$p$};
\filldraw (0,0) circle (2 pt);
\end{tikzpicture}
\quad $=-g(2\pi)^4\delta(p+q+k)$\vspace*{1pt}
\item Integrate over all momenta;
\item Divide by the symmetry factor.
\end{enumerate}

\begin{figure}[H]
\begin{center}
\begin{tikzpicture}
\draw[->-, very thick] (0,0) to (2,0);
\filldraw (2,0) circle (2 pt);
\draw (1,0) node[anchor=south] {$p$};
\draw (3.6,0) node[anchor=east]{$q$};
\draw (0,0) node[circle, draw= black, fill=black, inner sep =2.5pt] {};
\draw[->-, very thick] (2,0) arc(-180:180:0.5);
\draw[->-, very thick] (7,0) to (5,0);
\draw (6.1,0) node[anchor=south] {$p$};
\draw (5,0) node[circle, draw= black, fill=black, inner sep =2.5pt] {};
\draw[->-, very thick] (7,0) arc(-180:180:0.5);
\filldraw (7,0) circle (2 pt);
\draw (8.6,0) node[anchor=east]{$q$};
\end{tikzpicture}
\end{center}
\caption{\textit{Feynman diagrams related by switching the direction of a propagator are inequivalent.}}
\label{fig:inequivalent}
\end{figure}
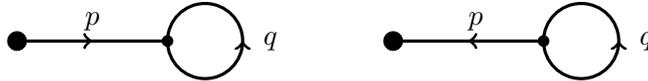

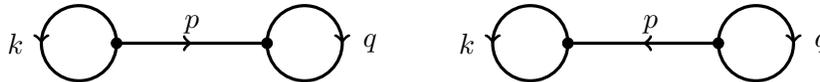
\begin{figure}[H]
\begin{center}
\begin{tikzpicture}[baseline={([yshift={0.1ex}]current bounding box.north)}]
\draw[->-, very thick] (0,0) to (2,0);
\filldraw (0,0) circle (2pt);
\filldraw (2,0) circle (2pt);
\draw[->-, very thick] (0,0) arc(0:360:0.5);
\draw (1,0) node[anchor=south] {$p$};
\draw (3.6,0) node[anchor=east]{$q$};
\draw (-1.1,0) node[anchor=east]{$k$};
\draw[->-, very thick] (2,0) arc(180:-180:0.5);
\draw[->-, very thick] (8,0) to (6,0);
\draw (7.1,0) node[anchor=south] {$p$};
\draw (4.9,0) node[anchor=east]{$k$};
\draw (9.6,0) node[anchor=east]{$q$};
\draw[->-, very thick] (6,0) arc(0:360:0.5);
\draw[->-, very thick] (8,0) arc(180:-180:0.5);
\filldraw (6,0) circle (2pt);
\filldraw (8,0) circle (2pt);
\end{tikzpicture}
\end{center}
\caption{\textit{Feynman diagrams related by relabelling of propagators are equivalent.}}
\label{fig:equivalent}
\end{figure}

\begin{figure}[H]
\begin{center}
\begin{tikzpicture}[baseline={([yshift={0.1ex}]current bounding box.north)}]
\filldraw (0,0) circle (2pt);
\filldraw (4,0) circle (2pt);
\draw[->-, very thick] (0,0) to (4,0);
\draw (2,0) node[anchor=south] {$p$};
\draw (2.2,1.3) node[anchor=east]{$q$};
\draw (2.2,-1.3) node[anchor=east]{$k$};
\draw[->-, very thick] (0,0) arc(180:0:2 and 1);
\draw[->-, very thick] (0,0) arc(-180:0:2 and 1);

\filldraw (11,0) circle (2pt);
\filldraw (7,0) circle (2pt);
\draw[->-, very thick] (7,0) to (11,0);
\draw (9,0) node[anchor=south] {$p$};
\draw (9.2,1.3) node[anchor=east]{$q$};
\draw (9.2,-1.3) node[anchor=east]{$k$};
\draw[->-,very thick] (11,0) arc(0:180:2 and 1);
\draw[->-,very thick] (7,0) arc(-180:0:2 and 1);
\end{tikzpicture}
\end{center}
\caption{\textit{Relabelling the propagators gives a residual symmetry factor of $3!$ for the left diagram and $2!$ for the right.}}
\label{fig:residual}
\end{figure}
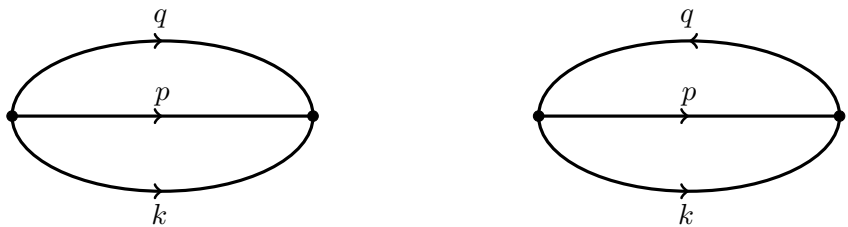

\indent We now consider how these rules are modified in the framework of relative locality introduce in the previous section. Let us begin with rule 4), integrate over all momenta. This is equivalent to introducing a measure on momentum space, call it $d\mu(p)$. For the time being we will make no assumptions about the measure other than demanding it reduce to the standard Lebesgue measure in the limit when momentum space becomes a linear manifold\footnote{An obvious choice would be $d\mu(p) = \sqrt{g(p)}d^4p$.}. Given $d\mu(p)$ we define $\delta(p,q)$ to be a delta function compatible with this measure, that is:
\begin{align}
\int d\mu(p) \delta(p,q) f(p) = f(q)
\end{align}
for any function $f:\mathcal{P} \to \mathcal{P}$. Note that this delta function is assumed to be symmetric upon interchange of its arguments, i.e. $\delta(p,q) = \delta (q,p)$. \\
\indent In deriving the original Feynman rules we tacitly assumed that the change of variables $p \to -p$ has unit Jacobian. In relative locality the equivalent change of variables is $p \to \om p$ which has Jacobian $\left|\det(d_p \om)\right| = \left|\det(I^p)\right|$.\footnote{Our assumption of a unique inverse is critical here; it is equivalent to demanding that $\om$ be invertible which in turn is necessary to even define this change of variables.} A priori this quantity could differ from unity which amounts to breaking the symmetry associated with flipping the direction of a propagator. Therefore, diagrams which are related by such a transformation should be regarded as inequivalent, see Figure \ref{fig:inequivalent}.
\noindent Diagrams do, however, still posses a symmetry under relabelling of propagators, for example the diagrams shown in Figure \ref{fig:equivalent} are equivalent.
All of this implies that we must propose a different interpretation of the symmetry factor, rule 5). A bit of thought suggests the following modification: Divide by $2^P$, where $P$ is the number of propagators appearing in the diagram, then divide by a factor associated with any residual symmetries of the diagram. The diagrams in Figures \ref{fig:inequivalent}, \ref{fig:equivalent} have no residual symmetry whereas those in Figure \ref{fig:residual} have residual symmetry factors of $3!$ and $2!$ respectively, given by relabelling the propagators. 
We turn next to Rule 1), the factor associated with the propagator.\footnote{In what follows we will drop all factors of $(2\pi)^4$.} The propagator must have a single simple pole at the particles mass which, given the definition of mass in Relative Locality, suggest that we make the following replacement:
\begin{equation}
p^2 + m^2 \to D^2(p) + m^2.
\end{equation}
where $D(p)$, we recall, is the distance of $p$ from the origin, as measured by the momentum space geometry $g(p)$.

\begin{figure}[H]
\begin{center}
\begin{tikzpicture}[baseline={([yshift={0.2ex}]current bounding box.north)}]
\filldraw (0,0) circle (2pt);
\filldraw (4,0) circle (2pt);
\filldraw (2,0) circle (2pt);
\draw[->-, very thick] (0,0) to (2,0);
\draw[->-, very thick] (2,0) to (4,0);
\draw (1,0) node[anchor=south] {$p$};
\draw (3,0) node[anchor=south] {$q$};
\end{tikzpicture}
\end{center}
\caption{\textit{Conserving momentum at a two point vertex. }}
\label{fig:twopoint}
\end{figure}
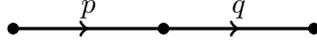

\indent Rule 2) requires no modification and so we come to rule 3), the factor assigned to a vertex. What properties should the modified factor posses? First, it should reduce to the original in the case where momentum space is a linear manifold. Second, it should respect the statistics of our particles. 
It is well known that in standard QFT scalar particles obey Bose statistics.
In our case since we modify the addition rule and relax the notion of locality,
we could also relax the bose statistics and investigate non trivial field statistics.
In this work we will take the simplest hypothesis and assume
 that we have  Bose statistics  in the present framework as well. Therefore, our factor must be symmetric upon interchange of momentum labels.  Given that the combination rule is neither associative nor commutative there are several choices we could make, we will consider three of them in detail. Assuming all particles are incoming to the vertex the first of these is:
\begin{equation}\label{delta1}
\begin{aligned}
\Delta_1 &= \frac{1}{6}\big[\delta(p \op (q\op k)) + \delta(p \op (k\op q)) + \delta(q \op (p\op k)) + \delta(q \op (k\op p))\\ 
&\qquad + \delta(k \op (p\op q)) + \delta(k \op (q\op p))\big],
\end{aligned}
\end{equation}
where we have used the simplified notation $\delta(p,0) = \delta(p)$. In this option we always assume that the second and third terms in the sum are grouped together.\footnote{Another, nearly equivalent, choice would be to group the first two terms together.}  The second choice includes all possible groupings and we write it as:
\begin{equation}\label{delta2}
\begin{aligned}
\Delta_2 &= \frac{1}{12}\sum_{\mathcal{K}(p,q,k)}\delta\left(\mathcal{K}(p,q,k)\right),
\end{aligned}
\end{equation}
where $\mathcal{K}(p,q,k)$ represents a possible ordering of momenta. The final option is similar to $\Delta_1$ but we move the grouped factors to the other side of the delta function, this gives
\begin{equation}\label{delta3}
\begin{aligned}
\Delta_3 &= \frac{1}{6}\big[\delta(p, \om (q\op k)) + \delta(p, \om (k\op q)) + \delta(q, \om (p\op k)) + \delta(q, \om (k\op p))\\ 
&\qquad + \delta(k, \om (p\op q)) + \delta(k, \om (q\op p))\big].
\end{aligned}
\end{equation}
\indent The difference between $\Delta_1$ and $\Delta_2$ is related to the discrepancy between $\delta(p\op q,0)$ and $\delta (q \op p,0)$ whereas the difference between $\Delta_1$ and $\Delta_3$ is related to the discrepancy between $\delta (p\op q,0)$ and $\delta (p,\om q)$. To gain some understanding of these discrepancies let us integrate these delta functions against an arbitrary function $f(p)$, we start with $\delta (p \op q)$:
\begin{align*}
\int d\mu(p) \delta(p\op q,0)f(p) &= \left\vert\det \left(V^{\om q}_{0}\right)\right\vert^{-1}f\left(\om q)\right).
\end{align*}
The calculation for $\delta (q \op p)$ is identical and yields:
\begin{align*}
\int d\mu(p) \delta(q\op p,0)f(p) &= \left\vert\det \left(U^{\om q}_{0}\right)\right\vert^{-1}f\left(\om q\right).
\end{align*}
Obviously these results would be interchanged if we had instead integrated over $q$. It remains to consider the value obtained from $\delta (p, \om q)$:
\begin{align*}
\int d\mu(p) \delta(p, \om q) f(p) &= f(\om q).
\end{align*}
Note that if we interchanged the roles of $p$ and $q$ in the previous integral we would obtain:
\begin{align*}
\int d\mu(p) \delta(q, \om p) f(p) &= \left\vert \det \left( I^{q}\right)\right\vert f(\om q).
\end{align*}
We see that the differences between the $\Delta_i$ is governed by the extent to which the determinant of the left or right transport operator differs from unity. \\
\indent It still remains to choose which $\Delta_i$ to use as a vertex factor. To motivate this choice let us imagine conserving momentum at a ``two point vertex'', see figure \ref{fig:twopoint}.  
\noindent Our prescription for conserving momentum should give $p=q$, i.e. $\int d\mu(q) \Delta_i(p,q) = 1$.  Both $\Delta_1$ and $\Delta_2$ yield a factor of
\begin{align*}
\frac{1}{2}\int d\mu(q)\left(\delta(p \om q) + \delta (\om q \op p)\right) = \frac{1}{2}\left|\det\left(I^p\right)\right|^{-1}\left(\left|\det\left(U^{\om p}_0\right)\right|^{-1} +  \left|\det\left(V^{\om p}_0\right)\right|^{-1}\right),
\end{align*}
whereas $\Delta_3$ gives
\begin{align*}
\int d\mu(q) \delta (p, q) = 1.
\end{align*}
This strongly suggests that we adopt $\Delta_3$ as our vertex factor and we will do so for the remainder of the paper. To keep notation simple we drop the $3$ and denote our vertex factor by  $-g\Delta(p,q,k)$.\\
\indent In summary, the modified generating functional is expanded as a sum of all  Feynamn diagrams with $E$ external points, $P$ propagators and $V$ vertices, where $E = 3v - 2P$. For each such diagram we include all possible orientations of propagator momenta that are inequivalent under relabelling. A numerical value is then assigned to these diagrams by means of the following Feynman rules:
\begin{enumerate}
\item To each propagator, \qquad  \begin{tikzpicture}[baseline={([yshift={-3.5ex}]current bounding box.north)}]
\draw[->-, very thick] (0,0) to (2,0);
\draw (1,0) node[anchor=south] {$p$};
\end{tikzpicture}
\quad $=\dfrac{i}{D^2(p) + m^2}$;
\item To each external point, 
\qquad  \begin{tikzpicture}[baseline={([yshift={-3.5ex}]current bounding box.north)}]
\draw[->-, very thick] (2,0) to (0,0);
\draw (1,0) node[anchor=south] {$p$};
\draw (0,0) node[circle, draw= black, fill=black, inner sep =2.5pt] {};
\end{tikzpicture}
\quad $=J(p)$;
\item To each vertex, \qquad  \begin{tikzpicture}[baseline={([yshift={-7ex}]current bounding box.north)}] 
\draw[->-, very thick] (-0.707,-0.707) to (0,0);
\draw[->-, very thick] (0.707,-0.707) to (0,0);
\draw[->-, very thick] (0,1) to (0,0);
\draw (-0.707,-0.707) node[anchor=south] {$q$};
\draw (0.707,-0.707) node[anchor=south] {$k$};
\draw (0,1) node[anchor=east] {$p$};
\filldraw (0,0) circle (2 pt);
\end{tikzpicture}
\quad $=-g\Delta(p,q,k)$\vspace*{1pt}
\item Integrate over all momenta using the measure $d\mu(p)$;
\item Divide by $2^P$ times the residual symmetry factor.
\end{enumerate}

\subsection{Modified Generating Functional and Action}
Having derived a set of Feynman rules we can now write down a generating functional for our theory. It is a straight forward exercise to see that the generating functional for $\ph^3$--theory in relative locality is given by:
\begin{equation}\label{defgen}
\begin{aligned}
Z_\mathrm{RL}(J) &\propto \exp\left(-\frac{g}{3!}\int d\mu(p)\int d\mu(q)\int d\mu(k) \Delta(p,q,k)\frac{\delta}{\delta J(p)}\frac{\delta}{\delta J(q)}\frac{\delta}{\delta J(k)}\right)\\
&\qquad \times \exp\left(\frac{i}{2}\int d\mu(p){J}(p)\left(D^2(p) + m^2\right)^{-1}{J}(\ominus p)\right),
\end{aligned}
\end{equation}
where the proportionality constant is fixed by demanding $Z_\mathrm{RL}(0) = 1$. The functional derivatives are defined to yield the delta function introduced in the previous section, viz
\begin{align}
\frac{\delta}{\delta{J}(p)}{J}(q) = \delta(p,q).
\end{align} 
To extract an action from this generating functional we need to evaluate the functional derivatives. This can be done by re--introducing scalar fields $\ph(p)$ as follows:
\begin{align*}
Z_\mathrm{RL}({J}) &\propto \exp\left(-\frac{g}{3!}\int d\mu(p)\int d\mu(q)\int d\mu(k) \Delta(p,q,k)\frac{\delta}{\delta {J}(p)}\frac{\delta}{\delta {J}(q)}\frac{\delta}{\delta {J}(k)}\right)\\
&\qquad \times \exp\left(\frac{i}{2}\int d\mu(p){J}(p)\left(D^2(p) + m^2\right)^{-1}{J}(\ominus p)\right)\\
&\qquad \times \int\mathcal{D}\ph \exp\left(-\frac{i}{2}\int d\mu(p)\left(D^2(p)+m^2\right){\ph}(p){\ph}(\ominus p)\right),
\end{align*}
where we have used that $Z_\mathrm{RL}$ is only defined up to a numerical factor. We can now bring the factor containing ${J}$ into the functional integral and then perform the change of variables ${\ph}(p) \to {\ph}(p) - {J}(p)(D^2(p) + m^2)^{-1}$. After some cancellation we find that the argument of the exponential in the path integral is given by
\begin{align*}
-\frac{i}{2}\int d\mu(p)\Big[&{\ph}(p){\ph}(\om p)\left(D^2(p)+m^2\right) - {J}(p){\ph}(\om p) - {\ph}(p){J}(\om p)\frac{D^2(p)+m^2}{D^2(\om p)+m^2} \\
&\quad + {J}(p){J}(\om p)\left(\left(D^2(\om p)+m^2\right)^{-1} - \left(D^2(p)+m^2\right)^{-1}\right)\Big].
\end{align*}
\indent The non--linear terms in ${J}$ will cancel if we demand $D^2(p) = D^2(\om p)$. This requirement is physically reasonable since $D^2(p)$ yields the squared mass of a particle with momentum $p$. On the other hand, $\om p$ simply represents a reversal in the direction of a particles momentum; it turns an incoming particle into an outgoing one and vice versa. This operation should not alter the mass of the particle and so $D^2(\om p) = -m^2 = D^2(p)$. The term quadratic in $J$ now drops out of the integrand and it becomes a simple matter to evaluate the functional derivatives appearing in \eqref{defgen}. In doing so we will make the assumption $\vert \det(I^{p})\vert = 1$ as assuming otherwise would make the result untenable. After we evaluate the functional derivatives we can read off the action as the argument of the exponential, we find
\begin{equation}
\begin{aligned}
S_\mathrm{RL} &= -\frac{1}{2}\int d\mu(p)\left(D^2(p)+m^2\right){\ph}(p){\ph}(\ominus p) \\
&\qquad + \frac{g}{3!}\int d\mu(p)\int d\mu(q)\int d\mu(k)\Delta(p,q,k){\ph}(\om p){\ph}(\om q){\ph}(\om k).
\end{aligned}
\end{equation}
The fields $\ph(p)$ commute and so the six terms in $\Delta(p,q,k)$ collapse to $\delta (p, \om (q\op k)$, which we can then eliminate by integrating over $p$ to obtain
\begin{equation}
\begin{aligned}
S_\mathrm{RL} &= -\frac{1}{2}\int d\mu(p)\left(D^2(p)+m^2\right){\ph}(p){\ph}(\ominus p) \\
&\qquad + \frac{g}{3!}\int d\mu(q)\int d\mu(k){\ph}(q \op k){\ph}(\om q){\ph}(\om k).
\end{aligned}
\end{equation}
\indent Finally we require that $S_\mathrm{RL}$ be real valued and so we impose the reality condition $\ph(\om p) = \ph^*(p)$; note though that for this prescription to work we also require 
\be
\om(p \op q) = (\om p) \op (\om q),\quad \mathrm{or}\quad
 \om(p\op q) = (\om q) \op (\om q).
 \ee 
 The first condition demands that $\ominus$ is a morphism while the second that it is an anti--morphism.
 These are the two conditions that respect the reality condition.
 Thus, the final form of our action is given by
\begin{equation}
\begin{aligned}\label{SRL}
S_\mathrm{RL} &= -\frac{1}{2}\int d\mu(p)\left(D^2(p)+m^2\right){\ph}(p){\ph}^*(p) \\
&\qquad + \frac{g}{3!}\int d\mu(q)\int d\mu(k){\ph}(q \op k){\ph}^*(q){\ph}^*(k).
\end{aligned}
\end{equation}
One key property of the action is its covariance under momentum space diffeomorphisms.
If one assumes that the integration measure is diffeomorphism invariant, i.e. $ d \mu (f(p)) = d \mu(p)$ for a diffeo
$f:\mathcal{P} \to \mathcal{P}$, that fixes the identity $f(0)=0$.
Then the Relative locality action satisfies 
\be
S_\mathrm{RL}(g,\oplus,\ph) = S_\mathrm{RL}(g_{f},\oplus_{f},\ph_{f})
\ee 
where 
\be
\ph_{f}(p) \equiv \ph(f(p)),\qquad p\oplus_{f}q \equiv f^{-1}(f(p)\oplus f(q)), 
\ee
while $g_{f}$ is the pull backed metric.

\section{Covariant Fourier Transform}
To explore the spacetime properties, in particular locality, of $S_\mathrm{RL}$ we need to compute its Fourier transform. Unfortunately we immediately run into a major impediment, the standard Fourier kernel $\exp(i p\cdot x)$ is not covariant and therefore its use would break the (momentum space) diffeomorphism covariance of our action. Instead we need to develop a generalization of the Fourier kernel which \emph{is} invariant under such diffeomorphisms. We begin by introducing Synge's world--function.

\subsection{Synge's World--Function}
\begin{figure}[H]
\begin{center}
\begin{tikzpicture}[baseline={([yshift={0.1ex}]current bounding box.north)}]
\filldraw (0,0) circle (2pt);
\filldraw (-4,2) circle (2pt);
\filldraw (1,2.5) circle (2pt);
\draw[->-, very thick] (0,0) to[out=130, in=-35] (-4,2);
\draw[->-, very thick] (0,0) to[out=30, in=190] (1,2.5);
\draw[->,red, thick] (0,0)to (-1.2856,1.5321);
\draw[->,blue, thick] (0,0)to (1.7321,1);
\draw [dotted, very thick] (-4,2) to[out=20, in=170] (1,2.5);
\draw (0,0) node[anchor=north] {$p'$};
\draw (-4,2) node[anchor=east]{$p$};
\draw (1,2.5) node[anchor=west]{$R_{p'}(p)$};
\draw (-1.2856,1.5321) node[anchor=south] {$\sig_{\mu'}(p,p')$};
\draw (1.7321,1) node[anchor=west] {$\sig^R_{\mu'}(p,p')$};
\end{tikzpicture}
\end{center}
\caption{\textit{Comparing $\sig_{\mu'}(p,p')$ and $\sig_{\mu'}^R(p,p')$. The thick black lines connecting $p'$ to $p$ and $R_{p'}(p)$ represent the unique geodesic interpolating between the two points.}}
\label{fig:sigvssigR}
\end{figure}
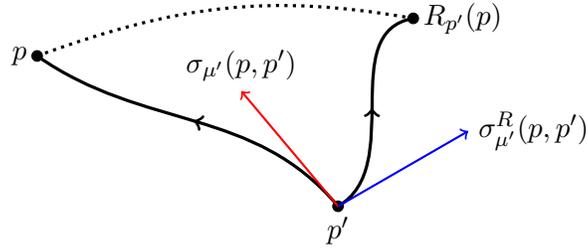

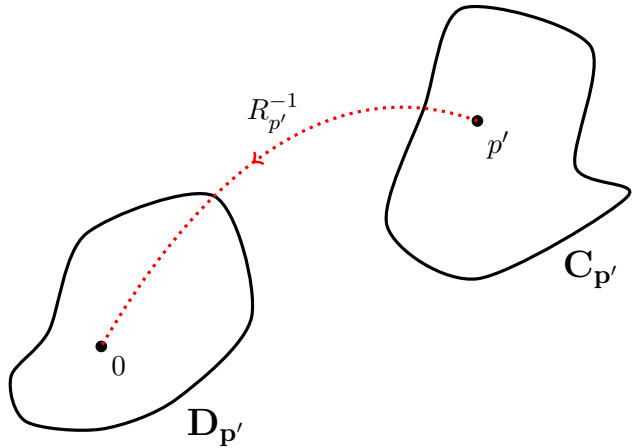
\begin{figure}[H]
\begin{center}
\begin{tikzpicture}[baseline={([yshift={0.1ex}]current bounding box.north)}]
\draw[very thick] plot [smooth cycle] coordinates {(0,0) (1,-0.1) (2,0.3) (3,1.4) (2.5,3) (0.8,2.5) (0.3,1.2) (-0.2,0.6) };
\draw[very thick] plot [smooth cycle] coordinates {(6,1.9) (8,3) (7.3,3.4) (7.5,5) (5.8,5.5) (5.3,4.2) (4.8,2.6) };
\filldraw (6,4) circle (2pt);
\filldraw (1,1) circle (2pt);
\draw [->-,dotted, very thick, red] (6,4) to[out =160, in=60] (1,1);
\draw (1,1) node[anchor=north west] {$0$};
\draw (6,4) node[anchor=north west] {$p'$};
\draw (2.8,4.5) node[anchor=north west] {$R_{p'}^{-1}$};
\draw (7,2.4) node[anchor=north west] {\large $\mathbf{C_{p'}}$};
\draw (2,0.3) node[anchor=north west] {\large $\mathbf{D_{p'}}$};
\end{tikzpicture}
\end{center}
\caption{\textit{The domain, $C_{p'}$, of $\sig(p,p')$ is mapped via $R_{p'}^{-1}$ to the domain, $D_{p'}$, of $\sig^R(p,p')$. }}
\label{fig:CvsD}
\end{figure}

The world--function was introduced by Synge (see \cite{SyngeGR}) in the context of General Relativity but the results apply equally well to a curved momentum space. Consider two points $p,p' \in \mathcal{P}$ and let $\gamma(\tau)$ be a geodesic connecting $p$ to $p'$ then the world--function, $\sig(p,p')$ is defined via
\begin{align}\label{wf}
\sig(p,p') = \frac{1}{2}\int_0^1 d\tau g^{\mu\nu}(\gamma(\tau))\frac{d\gamma_\mu(\tau)}{d\tau}\frac{d\gamma_\nu(\tau)}{d\tau}.
\end{align}
This integral is precisely the one used in deriving the geodesic equation, see Appendix \ref{ap:geodesics}, and so 
\begin{align}
\sig(p,p') = \frac{1}{2}D^2(p,p'), 
\end{align}
which implies that the world--function is half the square of the geodesic distance between $p$ and $p'$.\\
\indent Most properties of the world--function derived in \cite{SyngeGR} follow from the fact that the integrand appearing in \eqref{wf} is constant along a geodesic. As this condition holds for our definition of a geodesic, see Appendix \ref{ap:geodesics}, we can important these properties directly, the most important of which is the defining differential equation satisfied by $\sig$:
\begin{align}\label{wfde}
2\sig(p,p') = \sig_\mu(p,p')\sig^\mu(p,p') = \sig_{\mu'}(p,p') \sig^{\mu'}(p,p'),
\end{align}
where we have employed the standard notation
\begin{align*}
\nabla_{p_\mu} \sig(p,p') = \sig^\mu(p,p') \qquad \mathrm{and} \qquad \nabla_{p'_\mu} \sig(p,p') = \sig^{\mu'}(p,p').
\end{align*}
One can also examine the behaviour of the world--function (and its derivatives) as $p \to p'$ or vice versa. This is known as the ``coincidence limit'' and is indicated by square brackets, $[\ldots]$; e.g. $[\sig] = 0$. Besides this rather obvious one, the most common coincidence limits are given by 
\begin{align*}
[\sig^\mu] &= [\sig^{\mu'}] = 0\\
[\sig^{\mu\nu}] &= [\sig^{\mu'\nu'}] = -[\sig^{\mu\nu'}]= g^{\mu\nu}.
\end{align*}
The coincidence limit will not be of great important in this paper, but we refer the read to \cite{SyngeGR} for a complete discussion.\\
\indent The covariant derivatives of $\sig(p,p')$, being the derivatives of a bi--scalar, behave as contravariant vectors. In particular, $\sig_\mu(p,p')$ transforms as a scalar at $p'$ and a contravariant vector at $p$ and vice versa for $\sig_{\mu'}(p,p')$. Therefore, if $x^{\mu'} \in T^*_{p'}\mathcal{P}$ then $x^{\mu'}\sig_{\mu'}(p,p')$ transforms as a scalar at both $p$ and $p'$ and so a natural definition of the covariant Fourier kernel is $\exp(ix^{\mu'}\sig_{\mu'}(p,p'))$. This isn't quite right though. In the limit where the geometry of momentum space is trivial we have $\exp(ix^{\mu'}\sig_{\mu'}(p,p')) \to \exp(ix^{\mu}(p-p')_{\mu})$ and the dependence on $p'$ persists; an undesirable outcome. The solution is to introduce a translated version of the world--function and of its derivative at $p'$:
\be
\sig^R(p,p') \equiv \sig(R_{p'}(p), p'),\qquad \sig^{R \mu'}(p,p') \equiv \left(\nabla_{p'_\mu}\sig(p,p')\right)\Big\vert_{p = R_{p'}(p)}.
\ee
{We could have also defined a left translated version of the world--function, $\sig^L(p,p') \equiv \sig(L_{p'}(p),p')$, but we chose $\sig^R$ for the sake of definiteness.}
A graphical comparison of $\sig_{\mu'}(p,p')$ and $\sig^R_{\mu'}(p,p')$ is given in Figure \ref{fig:sigvssigR}. The kernel $\exp(ix^{\mu'}\sig^R_{\mu'}(p,p'))$ is then covariant and reduces to $\exp(ix^\mu p_\mu)$ in the limit of flat momentum space. It will form the basis for defining the covariant Fourier transform .
\indent Before we continue there are some technical issues regarding the domain of the world--function which need to be discussed. Fix the point $p' \in \mc{P}$. The definition of $\sig(p,p')$ assumes the existence of a \textit{unique} geodesic connecting $p$ to $p'$; a condition which is not, in general, satisfied for two arbitrary points in $\mc{P}$. To ensure the world--function remains single valued we need to restrict its domain to a ``normal convex neighbourhood'' (see \cite{ConvexNormal}) of $p'$, denoted $C_{p'}$. More specifically, $C_{p'}$ is a subset of $\mc{P}$ containing $p'$ such that, given another point $p \in C_{p'}$ there exists a unique geodesic, completely contained in $C_{p'}$, connecting $p'$ and $p$.\footnote{The existence of such a neighbourhood for any $p' \in \mc{P}$ is guaranteed by Whiteheads theorem \cite{ConvexNormal}.} Our primary interest, however, is in the translated world--function $\sig^R(p,p')$ which will have a domain of definition given by $D_{p'} = R_{p'}^{-1}(C_{p'})$.
It is important to note that even if this domain depends on $p'$ it is always a domain centered around the identity,
i.e.  $0 \in D_{p'}$. See Figure \ref{fig:CvsD}.

\subsection{Van--Vleck Morette Determinant}
In this section we introduce the Van--Vleck Morette determinant (\cite{VanVleck},\cite{Morette},\cite{Poisson}), a quantity which will play an important role in our definition of the covariant Fourier transform. The change of variables $p_\mu \to Q'^{\mu} = \sig^{R\mu'}(p,p')$, where $Q'\in T_{p'}^{*}{\cal P}$ and $g^{-1}Q'\in T_{p'}{\cal P}$ is the initial velocity vector of the geodesic going from $p'$ to $p$. It  has Jacobian given by
\begin{align*}
d^4Q' 
= \left\vert\det\left( \sig^{R\mu\nu'}(p,p')\right)\right\vert d^4p,
\end{align*}
where  we have employed the notation
\begin{align*}
 \sig^{R\mu\nu'}(p,p') = \nabla^\mu \sig^{R \nu'}(p,p').
\end{align*} 
The Van--Vleck Morette determinant is the bi--scalar obtained from this Jacobian through multiplication by the metric determinant, in particular
\begin{align}\label{vanVleck}
\mc{V}(p,p') \equiv \frac{\left\vert\det\left( \sig^{R\mu\nu'}(p,p')\right)\right\vert}{\sqrt{g}_{p'}\sqrt{g}_{R_{p'}(p)}}.
\end{align}
It appears naturally in the  symplectic measure when we go from the symplectic coordinates $(Q',p')$ to the end point coordinates $(p,p')$ as
\be
\rd Q' \wedge \rd p' = \mc{V}(p,p') (\sqrt{g}_{p'}\rd^{4} p') \wedge (\sqrt{g}_{R_{p'}(p)} \rd^{4}p).
\ee
Note that the change of coordinates $ Q' \to p=R_{p'}^{-1}\left(\exp_{p'}(g^{-1}Q') \right) $ from $T^{*}_{p'}\mc{P}$ to
${\cal P}$,   is  the translated exponential map.
And the inverse Van--Vleck Morette determinant is the Jacobian for this transformation:
\begin{align}\label{changecor}
(\sqrt{g}_{R_{p'}(p)} \rd^4p) = \mc{V}^{-1}(p,p')\left(\frac{\rd^4Q'}{\sqrt{g}_{p'}}\right),
\end{align}
which highlights an important property of the Van--Vleck Morette determinant. If $p\in \mc{P}$ is such that $\mc{V}^{-1}(p,p') = 0$ then a change in $Q'$ produces no change in $p$ which is equivalent to making a change in the geodesic emanating from $p'$ but no change in the point at which the geodesic terminates; i.e. $p$ is a caustic. The reverse situation, where $\mc{V}(p,p') = 0$, is impossible since one cannot change the terminating point of a geodesic without altering the geodesics tangent vector at the sourcing point. Therefore, while the Van--Vleck Morette determinant is non--zero for all $p \in \mc{P}$ it does diverge at caustics. As a final note we observe that $\mc{V}(p,p')$ satisfies
\begin{align}
\mc{V}(0,p') = 1.
\end{align}

\subsection{Covariant Fourier Transform}
Heuristically, we expect the covariant Fourier transform to take functions on $\mathcal{P}$ and map them to functions on $T^*_{p'}\mathcal{P}$. It is natural then to introduce the notation 
\be
\mathcal{M}_{p'} \equiv T^*_{p'}\mathcal{P},
\ee which express that the cotangent plane at $p'$ acts as a ``spacetime'' at $p'$ for the Fourier transform. To formalize this initial expectation we fix a point $p' \in \mc{P}$ and choose a normal convex neighbourhood $C_{p'}$ giving $D_{p'} \equiv R_{p'}^{-1}(C_{p'})$ as the domain of $\sig^R(p,p')$. The measure on momentum space, denoted $d\mu(p)$ above, and on the dual spacetime are defined by
\begin{align*}
d{\mu}_{p'}(p) &= \sqrt{g}_{R_{p'}(p)} \rd^4p,\\
d\nu_{p'}(x) &= g^{-1/2}_{p'}\rd^4x,
\end{align*}
respectively. Let $\mc{L}^2_{{\mu}_{p'}}(D_{p'})$ denote the space of all functions on $\mc{P}$ which are square integrable with respect to $d {\mu}_{p'}$ and vanish outside of $D_{p'}$. The covariant Fourier transform (see \cite{covariantfourier, covariantfourier2} for earlier implementation of this object in a different context) is then the map, $\mc{F}_{p'}$, given by
\begin{align*}
\mathcal{F}_{p'}: \mathcal{L}^2_{ {\mu}_{p'}}(D_{p'}) &\to \mathcal{L}_{\nu_{p'}}^2(\mathcal{M}_{p'})\\
f(p) &\mapsto \hat{f}_{p'}(x),
\end{align*}
where
\begin{align}\label{FT}
\hat{f}_{p'}(x) \equiv \int_{D_{p'}} d\mu_{p'}(p) \mathcal{V}^{1/2}(p,p')\exp\left(-ix^{\mu'}\sig^R_{\mu'}(p,p')\right)f(p).
\end{align}
Unless $D_{p'}=\mc{P}$, the covariant Fourier transform is not surjective and therefore is not invertible on all of $\mc{L}^2_{\nu_{p'}}(\mc{M}_{p'})$. This difficulty can be circumvented by restricting to the image of $\mc{F}_{p'}$, i.e. $\hat{f}_{p'}(x) \in \mc{F}_{p'}(\mc{L}_{\hat{\mu}_{p'}}(D_{p'}))$, which allows us to define the inverse Fourier transform as
\begin{align}\label{IFT}
\mathcal{F}_{p'}^{-1}(\hat{f}_{p'})(p) \equiv \int_{\mc{M}_{p'}} d\nu_{p'}(x)\mathcal{V}^{1/2}(p,p') \exp\left(ix^{\mu'}\sig^R_{\mu'}(p,p')\right)\hat{f}_{p'}(x),
\end{align}
for $p \in D_{p'}$ and zero otherwise. The Fourier transform of a function $\hat{f}_{p'}(x) = \mc{F}_{p'}(f(p))(x)$, one will notice, depends on the choice of base point $p'$. One does not, therefore, obtain a single Fourier transform but rather a continuum as the base point $p'$ varies throughout $\mc{P}$.\\
\indent As an initial application of this formalism we will consider the Fourier representation of $\delta(p,q)$, the delta function on $\mc{P}$. Assuming $p,q \in D_{p'}$ we posit
\begin{align}\label{deltaP}
\delta(p,q) \equiv \int d\nu_{p'}(x) \mathcal{V}^{1/2}(p,p')\mathcal{V}^{1/2}(q,p') \exp\left[ix^{\mu'}\left(\sig^R_{\mu'}(p,p') - \sig^R_{\mu'}(q,p')\right)\right].
\end{align}
This formula is explicitly verified in Appendix \ref{ap:technical} but we note here that the proof depends crucially on the fact, left implicit in the above formula, that the integral is taken over all of $\mc{M}_{p'}$. One can also define a Fourier representation of the delta function on $\mc{M}_{p'}$, denoted $\delta_{p'}(x,y)$, by putting
\begin{align}\label{deltaTP}
\delta_{p'}(x,y)& = \int_{D_{p'}} d\mu(p) \mathcal{V}(p,p')\exp\left[i\sig^R_{\mu'}(p,p')\left(x^{\mu'} - y^{\mu'}\right)\right].
\end{align}
It is important to note this representation is not the usual delta function unless $D_{p'} = \mc{P}$. It is  a projector under convolution, that is 
\begin{equation}
\delta_{p'}(x,y) = \int_{{\cal M}_{p'}} d\nu_{p'}(z) \delta_{p'}(x,z)\delta_{p'}(z,y).
\end{equation}
It therefore acts as an identity on the image of the Fourier transform i.e. on  $\mc{F}_{p'}(\mc{L}_{\hat{\mu}_{p'}}(D_{p'}))$. These properties are shown in Appendix \ref{ap:technical}. Note that a mathematical study of
a generalized Fourier transformation has already been done in \cite{MajidL} in the context of non--commutative $\mathrm{SU}(2)$ field theory.

\subsection{Plane waves}
In this section we introduce the notion of plane waves which turn out to be an efficient method for representing the covaiant Fourier transform. Formally, we define a plane wave, based at the point $p' \in \mathcal{P}$, to be the function of $p \in D_{p'}$ and $x \in \mathcal{M}_{p'}$ given by   
\begin{align}
e_{p'}(p,x) = \mathcal{V}^{1/2}(R_{p'}(p),p')\exp\left(-ix^{\mu'}\sig^R_{\mu'}(p,p')\right).
\end{align}
Recalling the defining differential equation for the world--function, equation \eqref{wfde}, a simple calculation shows that $e_{p'}(p,x)$ is an eigenfunction of the Laplacian on $\mc{M}_{p'}$,
\begin{align*}
g^{\mu'\nu'}(p')\frac{\partial}{\partial x^{\mu'}}\frac{\partial}{\partial x^{\nu'}}e_{p'}(p,x) &= -g^{\mu'\nu'}(p')\sig^R_{\mu'}(p,p')\sig_{\nu'}(p,p') e_{p'}(p,x)\\
&= -2\sig^R(p,p')e_{p'}(p,x)\\
&=-D^2(R_{p'}(p),p')e_{p'}(p,x).
\end{align*}
In particular, putting $p'=0$ we find  
\begin{align}\label{box}
D^2(p)e_{0}(p,x) = -\Box_x e_{0}(p,x);
\end{align} 
a result which will be important in the sequel since it is $D^2(p)$ which appears in the action, $S_{\mathrm{RL}}$. Returning to the definition of $e_{p'}(p,x)$ we see that the covariant Fourier transform, its inverse and the delta functions introduced in the previous section can be re--written as
\begin{align}
\hat{f}_{p'}(x) &= \int_{D_{p'}} d\mu_{p'}(p) e_{p'}(p,x)f(p),\label{planef}\\
f(p) &= \int_{\mc{M}_{p'}} d\nu_{p'}(x) e_{p'}^*(p,x)\hat{f}_{p'}(x),\label{planeif}\\
\delta(p,q) &= \int_{\mc{M}_{p'}} d\nu_{p'}(x) e_{p'}(p,x) e_{p'}^*(q,x),\\
\delta_{p'}(x,y) &= \int_{D_{p'}} d\mu_{p'}(p)e^*_{p'}(p,x)e_{p'}(p,y).
\end{align} 
The advantage of this notation becomes apparent when we attempt to prove the Plancherel formula, which states that 
\begin{align}
\int_{\mc{M}_{p'}} d\nu_{p'}(x) \hat{f}_{p'}(x) \hat{f}_{p'}^*(x) = \int_{D_{p'}} d\mu_{p'}(p) f(p)f^*(p),
\end{align}
provided $\delta_{p'}\circ \hat{f}_{p'} =\hat{f}_{p'}$, which ensures that $\hat{f}_{p'}$ is in the image of the Fourier transform.
The proof proceeds as follows, let $\hat{f}_{p'}(x) \in \mc{F}_{p'}(\mc{L}_{\hat{\mu}_{p'}}(D_{p'}))$ then
\begin{align*}
\int d\nu_{p'}(x) \hat{f}_{p'}(x) \hat{f}_{p'}^*(x)  &= \int d\nu_{p'}(x)d\mu_{p'}(p)d\mu_{p'}(q)e_{p'}(p,x)e^*_{p'}(q,x)f(p)f^*(q)\\
&=\int d\mu_{p'}(p)d\mu_{p'}(q)\delta(p,q)f(p)f^*(q)\\
&=\int d\mu_{p'}(p)f(p)f^*(p),
\end{align*}
which is the desired result. A similarly straightforward calculation will also verify our claim that \eqref{IFT} represents the inverse of $\mc{F}_{p'}$.\\
\indent Observe that the Fourier transform of a function lives in a particular cotangent space designated by $p'$. To understand the relationship between different choices of $p'$ we define a transport operator $T_{p',q'}(x,y)$ which satisfies
\begin{align}\label{T1}
\hat{f}_{p'}(x) \equiv \int_{\mc{M}_{q'}} d\nu_{q'}(y) T_{p',q'}(x,y) \hat{f}_{q'}(y).
\end{align} 
In other words, $T_{p',q'}$ maps the Fourier transform in one cotangent space to the Fourier transform in another. We can derive an explicit expression for the transport operator by taking the transform of a particular function twice, i.e. 
\begin{align*}
\hat{f}_{p'}(x) &= \int_{D_{p'}} d\mu_{p'}(p)e_{p'}(p,x)f(p) \\
&= \int_{D_{p'}\cap D_{q'}} d\mu_{p'}(p)\int_{\mc{M}_{q'}} d\nu_{q'}(y) e_{p'}(p,x)e^*_{q'}(p,y)\hat{f}_{q'}(x).
\end{align*}
In the second line we took the Fourier transform at $q'$ which requires $f(p)$ to vanish outside $D_{q'}$ and so we obtain the stated domain of integration $D_{p'}\cap D_{q'}$. Comparison with the definition of $T_{p',q'}$ in \eqref{T1} yields
\begin{align}
T_{p',q'} (x,y) = \int_{D_{p'}\cap D_{q'}} d\mu_{p'}(p) e_{p'}(p,x)e^*_{q'}(p,y).
\end{align}
In the limit where $p' = q'$ this transport operator is simply the delta function $\delta_{p'}(y,x)$, in all other cases $T_{p',q'}$ is a non--local operator.

\subsection{Star Product}
As a final piece of machinery we define a star product on $\mc{F}_{p'}(\mathcal{L}^2_{\hat{\mu}_{p'}}(D_{p'}))$ as follows
\begin{align*}
(\hat{f}_{p'} \star_{p'} \hat{g}_{p'})(x) \equiv \int_{\mc{M}_{p'}\times \mc{M}_{p'}} d\nu_{p'}(y)d\nu_{p'}(z) \omega_{p'}(x,y,z)\hat{f}_{p'}(y)\hat{g}_{p'}(z),
\end{align*}
where the kernel $\omega_{p'}(x,y,z)$ is given by
\begin{align}
\omega_{p'}(x,y,z) \equiv \int_{D_{p'}\times D_{p'}} d\mu_{p'}(p)d\mu_{p'}(q) e_{p'}(p\op q,x) e^*_{p'}(p,y) e^*_{p'}(q,z).
\end{align}
Note that the star product is defined only on functions living in the same  cotangent spaces $\mc{M}_{p'}=T_{p'}^*\mathcal{P}$. Let's take a moment to explore some of the properties this product possesses. First, the product of two plane waves yields the rather pleasing result (see \cite{FL1,FL2} for similar properties in quantum gravity)
\begin{align*}
e_{p'}(p,x) \star_{p'} e_{p'}(q,x)= e_{p'}(p \op q,x).
\end{align*}
Second, explicitly computing the star product of two functions, $(\hat{f}_{p'} \star_{p'} \hat{g}_{p'})(x)$, we find
\begin{align}\label{conv2}
\left(\hat{f}_{p'} \star_{p'} \hat{g}_{p'}\right)(x) &= \int d\mu_{p'}(p)d\mu_{p'}(q) e_{p'}(p\op q,x) f(p)g(q),
\end{align}
where $f(p)$ and $g(p)$ have Fourier transforms $\hat{f}_{p'}$ and $\hat{g}_{p'}$ respectively. Furthermore, since $\op$ is not commutative we can see that $\star_{p'}$ will also fail to commute. Finally, taking the convolution product of three functions
\begin{equation}\label{conv3}
\left(\hat{f}_{p'}\star_{p'}\left(\hat{g}_{p'}\star_{p'}\hat{h}_{p'}\right)\right)(x) =\int d\mu_{p'}(p)d\mu_{p'}(q)d\mu_{p'}(k)e_{p'}(p\op(q\op k),x) f(p) g(q)h(k),
\end{equation}
which demonstrates that the failure of $\op$ to associate propagates a similar failure into $\star_{p'}$.\\
\indent Let us now investigate the relationship between the star product and the standard point--wise product. Noting that $e_{p'}(0,x) = 1$ we can integrate \eqref{conv2} over $x$ to find
\begin{align}\label{intconv2}
\int d\nu_{p'}(x)\left(\hat{f}_{p'} \star_{p'} \hat{g}_{p'}\right)(x) &= \int d\mu_{p'}(p) \left\vert\det\left(V^{\om p}_{0}\right)\right\vert^{-1} f\left(\om p\right)g\left(p\right) 
\end{align}
On the other hand, if we compute the integral over the point--wise product $f_{p'}(x)g_{p'}^*(x)$ the Plancherel theorem will give the same result, less the factor of $\det(V)$. By setting $\vert \det(V^p_0)\vert = 1$ for all $p \in \mathcal{P}$ it follows that (the integral of) the star product and point--wise product match.\footnote{By virtue of \eqref{IUV} it follows that $\vert\det(U^p_0)\vert =1 $ for all $p \in \mathcal{P}$ as well.} In this sense, we can say the star product of two functions is a local object. Performing a similar computation for the star product of three functions we find
\begin{equation}\label{intconv3}
\begin{aligned}
\int d\nu_{p'}(x) \left(\hat{f}_{p'}\star_{p'}\left(\hat{g}_{p'}\star_{p'}\hat{h}_{p'}\right)\right)(x) =\int d\mu_{p'}(p)d\mu_{p'}(q)f(p\op q) g(\om p)h\left(\om q\right),
\end{aligned}
\end{equation}
where we have also made the change of variables $p,q \to \om p, \om q$. A bit of thought should convince the reader that \eqref{intconv3} bears little relation to the integral over the point--wise product of three functions, implying that the star product of three functions is a non--local object. This concludes the technical developments and we are now prepared to apply our formalism to the action $S_{\mathrm{RL}}$.

\subsection{Action in Spacetime}
For ease of notation we will not explicitly display the domain of integration in any integrals occurring in this section. Recall that $S_{RL}$, the momentum space action for our scalar field theory, is given by
\begin{equation}
\begin{aligned}
S_\mathrm{RL} &= -\frac{1}{2}\int d\mu_{p'}(p)\left(D^2(p)+m^2\right){\ph}(p){\ph}^*(p) \\
&\qquad + \frac{g}{3!}\int d\mu_{p'}(q)\int d\mu_{p'}(k){\ph}(q \op k){\ph}^*(q){\ph}^*(k).
\end{aligned}
\end{equation}
 Comparing the terms appearing above with equations \eqref{intconv2} and \eqref{intconv3}, and recalling that $\ph(\om p) = \ph^*(p)$, we can make the following replacements
\begin{align}\label{m2}
m^2\int d\mu_{p'}(p){\ph}(p){\ph}^*(p) = m^2\int d\nu_{p'}(x)\left(\hat{\ph}_{p'} \star_{p'} \hat{\ph}_{p'}\right)(x),
\end{align}\\
and
\begin{align}\label{g}
\int d\mu_{p'}(q)d\mu_{p'}(k){\ph}(q \op k){\ph}^*(q){\ph}^*(k) = \int d\nu_{p'}(x) \left(\hat{\ph}_{p'}\star_{p'}\left(\hat{\ph}_{p'}\star_{p'}\hat{\ph}_{p'}\right)\right)(x).
\end{align}
 As discussed in the previous section the integral appearing in equation \eqref{m2} is local whereas the one appearing in equation \eqref{g} is not. \\
\indent The $D^2(p)$ term is more complex and we can not make the simple replacements used above. We proceed by taking the covariant Fourier transform of $\ph(p)$ and $\ph^*(p)$
\begin{align}\label{DS1}
\int d\mu_{p'}(p) D^2(p) \ph(p)\ph^*(p) &= \int d\mu_{p'}(p)d\nu_{p'}(x)d\nu_{p'}(y) D^2(p) e_{p'}^*(p,x)e_{p'}(p,y)\hat{\ph}_{p'}(x)\hat{\ph}^*_{p'}(y).
\end{align}
To proceed we would like to use equation \eqref{box} and exchange $D^2(p)$ for derivatives of a plane wave, but doing so requires a plane wave based at $p' =0$. As such we shift $e_{p'}(p,y)$ to $e_0(p,z)$ by introducing the translation operator $T_{p',0}(y,z)$, viz
\begin{align*}
D^2(p)e_{p'}(p,y) = \int d\nu_{0}(z) D^2(p)T_{p',0}(y,z)e_{0}(p,z) = -\int d\nu_{0}(z)T_{p',0}(y,z)\Box_z e_0(p,z)
\end{align*}
Integrating by parts moves the derivatives onto $T_{p',0}$ which allows us to translate the plane wave back to $p'$ by introducing another translation operator
\begin{align}
D^2(p)e_{p'}(p,y) &= -\int d\nu_{0}(z)d\nu_{p'}(a) e_{p'}(p,a)T_{0,p'}(z,a)\Box_z T_{p',0}(y,z).
\end{align}
We can now substitute this back into \eqref{DS1} and integrate over $p$ to obtain the delta function $\delta_{p'}(a,x)$, an integration over $a$ then gives
\begin{align*}
\int d\mu_{p'}(p) D^2(p) \ph(p)\ph^*(p)&= -\int d\nu_{p'}(x)d\nu_{p'}(y)d\nu_0(z) T_{0,p'}(z,x)\Box_z T_{p',0}(y,z)   \hat{\ph}_{p'}(x)\hat{\ph}^*_{p'}(y)\\
&=-\int d\nu_{p'}(y)d\nu_0(z) \left(\Box_z T_{p',0}(y,z)\right)\hat{\ph}_{0}(z)\hat{\ph}^*_{p'}(y)\\
&=-\int d\nu_{p'}(y)d\nu_0(z)  T_{p',0}(y,z) \Box_z\hat{\ph}_{0}(z)\hat{\ph}^*_{p'}(y).
\end{align*}
In the special case $p' = 0$ the translation operator becomes a delta function and integrating over $z$ we obtain the expected (and local) result $-\int d\nu_0(y)\hat{\ph}_0^*(y)\Box_y \hat{\ph}_{0}(y)$. On the other hand, if $p' \neq 0$ the transport operator will be de--localized and the overall result non--local. For ease of notation we will denote $(\Box_y \hat{\ph})_{p'}(y) = \int d\nu_0(z)  T_{p',0}(y,z) \Box_z\hat{\ph}_{0}(z)$ and so the $D^2(p)$ term can be written as
\begin{align}
\int d\mu_{p'}(p) D^2(p) \ph(p)\ph^*(p) &= -\int d\nu_{p'}(x) \left(\hat{\ph}_{p'}\star_{p'} (\Box \hat{\ph})_{p'}\right)(x),
\end{align}
recalling that the integral over the point--wise product of two functions is identical to the integral over the star product of two functions.\\
\indent Putting the results of this section together we find that the action for our scalar field theory, in the spacetime $\mc{M}_{p'}$, is given by
\begin{align}
S_{RL}^{p'} &=\frac{1}{2}\int d\nu_{p'}(x) \left[\left(\hat{\ph}_{p'}\star_{p'} (\Box \hat{\ph})_{p'}\right)(x) - m^2\left(\hat{\ph}_{p'} \star_{p'} \hat{\ph}_{p'}\right)(x)\right] \\
&\qquad + \frac{g}{3!}\int d\nu_{p'}(x) \left(\hat{\ph}_{p'}\star_{p'}\left(\hat{\ph}_{p'}\star_{p'}\hat{\ph}_{p'}\right)\right)(x).
\end{align}
Observe that the interaction term is non--local for any choice of $p'$ and the $m^2$ term is local for any choice of $p'$. The kinetic term on the other hand is local for $p' = 0$ but non--local for any other choice of the base point. 
This shows that if we denote $\hat{\ph}\equiv \hat{\ph}_{0}$, $d\nu(x)\equiv d\nu_{0}(x)$ and $\star\equiv \star_{0}$, the relative locality action becomes, simply
\begin{align}
S_{RL} &=\frac{1}{2}\int d\nu(x) \left[\left(\hat{\ph} \Box \hat{\ph}\right)(x) - m^2 \hat{\ph} \hat{\ph} (x)\right] + \frac{g}{3!}\int d\nu(x) \left(\hat{\ph} \star \left(\hat{\ph} \star \hat{\ph}\right)\right)(x).
\end{align}

\section{Conclusion}
Starting from the generating functional for standard $\ph^3$--theory we wrote down the corresponding momentum space Feynman rules which were then deformed to incorporate the non--linear structure of momentum space. We then derived the modified generating functional from which we were able to extract the action for our theory. A method for implementing a covariant Fourier transform was then developed along with a notion of plane waves and a star product. We found that the Fourier transform of a function on momentum space depended, implicitly, on a fixed point $p'$ in momentum space. Different choices of fixed point yielded different Fourier transforms with two such transforms being related by a non--local translation operator.\\
\indent Having developed this formalism in detail we used it to Fourier transform our action into spacetime. The resulting action depended, of course, on the choice of fixed point $p'$. The $m^2$ term in the action was found to be local for all choices of $p'$ and the interaction term non--local for all choices of $p'$. The kinetic term, however, was found to be local for $p' =0$ and non--local for all other choices of $p'$.\\
\indent This paper represents the first step towards developing quantum field theory in curved momentum space. To make phenomenological predictions though we need to incorporate fermions and gauge bosons into this framework, a task which will be the focus of future research.

\section*{Acknowledgement}
We would like to thank G. Amelino-Camelia and the quantum gravity group at PI for feedback on a talk given on this subject.
This research was supported in part by Perimeter Institute for Theoretical Physics. Research at Perimeter Institute is supported by the Government of Canada through Industry Canada and by the Province of Ontario through the Ministry of Research and Innovation. This research was also partly supported by grants from NSERC.

\appendix
\section{Geodesics}\label{ap:geodesics}
A geodesic can be defined as a path, $p(\tau)$, which parallel transports its own tangent vector. This requires $\dot{p}_\alpha \nabla^\alpha \dot{p}_\mu = 0$ and so the geodesic equation is given by:
\begin{align}
\frac{d^2p_\mu}{d\tau^2} + \Gamma^{\alpha\beta}_\mu\frac{dp_\alpha}{d\tau}\frac{dp_\beta}{d\tau} = 0\label{RLgeo}.
\end{align}
Alternatively, we can define a geodesic as a path which extremizes the distance between two points on the manifold. In general relativity, where the connection is metric compatible, these definitions are equivalent. This is not the case in relative locality where the connection is derived, not from a metric, but from the addition rule $\op$. In choosing between these definitions we note that the distance function $D_\gamma(p_0,p_1)$ is tied to the notion of mass and features prominently in the structure of relative locality. As such, it is natural to have a definition of geodesic which extremizes $D_\gamma$, and so we make this choice. We will now present a detailed derivation of the geodesic equation and explore some of its properties. \\
\indent Following the argument given in \cite{SyngeGR}, suppose we have two points $P,Q \in \mathcal{P}$ and an infinity of curves, $p_\mu(u,v)$ interpolating between $P$ and $Q$. The parameter $v$ indicates which curve is being considered while $u$ parametrizes the selected curve. We assume that $u$ varies between $u_0$ and $u_1$ so that $P,Q$ have coordinates $p_\mu(u_0,v)$ and $p_\mu(u_1,v)$ respectively. A geodesic is then a curve which gives a stationary value to the following integral for variations which leave the endpoints fixed\footnote{Such a curve will also give a stationary value to $D_\gamma$ so we are justified in considering the simpler function $I(v)$.}:
\begin{align}
I(v) = \frac{1}{2}\int_{u_0}^{u_1} g^{\mu\nu}\frac{dp_\mu}{du}\frac{dp_\nu}{du}du.
\end{align}
Introduce the tangent vectors $U_\mu = \partial p_\mu/\partial u$ and $V_\mu = \partial p_\mu/\partial u$, where $V_\mu$ vanishes at $u = u_0,u_1$. We then define the covariant derivative along the path $p_\mu$ by 
\begin{align}\label{DuDv}
\frac{D A_\mu}{du} = \frac{d A_\mu}{du} + \Gamma^{\alpha\beta}_\mu A_\alpha U_\beta \qquad \mathrm{and} \qquad \frac{D A_\mu}{dv} = \frac{d A_\mu}{dv} + \Gamma^{\alpha\beta}_\mu A_\alpha V_\beta,
\end{align}
where these definitions are extended to arbitrary tensors in the standard way. 
A brief calculation shows that $DU_\mu/dv = DV_\mu/du$, which we will make use of shortly. Demanding that $I(v)$ be stationary under variations which leave the end--points fixed is equivalent to the condition: $dI(v)/dv = 0$ for $V_\mu$ arbitrary, except at the end--points. Thus we proceed by differentiating $I(v)$, making use of the fact that $d/dv$ and $D/dv$ are interchangeable when applied to a scalar:
\begin{align}
\frac{d I(v)}{dv} &= \frac{1}{2}(u_1 - u_0)\int_{u_0}^{u_1} \left(\nabla^\rho g^{\mu\nu}V_\rho U_\mu U_\nu + 2g^{\mu\nu}U_\nu \frac{D U_\mu}{dv}\right)du\\
&=\frac{1}{2}(u_1 - u_0)\int_{u_0}^{u_1} \left(\left[N^{\rho\mu\nu}-2N^{\mu\rho\nu}\right]V_\rho U_\mu U_\nu - 2g^{\mu\nu}V_\mu \frac{D U_\nu}{du}\right)du.
\end{align}
Setting this to zero and expanding $DU_\nu/du$ using \eqref{DuDv} we find the geodesic equation:
\begin{align}\label{geo1}
\frac{dU_\alpha}{du} = \frac{1}{2}g_{\rho\alpha}\left[N^{\rho\mu\nu}-2N^{\mu\rho\nu}\right]U_\mu U_\nu - \Gamma^{\mu\nu}_\alpha U_\mu U_\nu.
\end{align}
This result can be simplified using equation \eqref{nonmetricity} which gives
\begin{align*}
\left[N^{\rho\mu\nu}-2N^{\mu\rho\nu}\right]U_\mu U_\nu 
&=2\left[T^{\rho\mu\nu} + \mathcal{N}^{\mu\nu}_\alpha g^{\alpha \rho}\right]U_\mu U_\nu.
\end{align*}
Substituting this back into \eqref{geo1}, noting that $\Gamma^{\mu\nu}_\rho U_\mu U_\nu = \Gamma^{(\mu\nu)}_\rho U_\mu U_\nu$ and using $\mathcal{N}^{\mu\nu}_\alpha = \Gamma^{(\mu\nu)}_\alpha - \left\lbrace \begin{smallmatrix} \mu\;\nu\\ \alpha \end{smallmatrix}\right\rbrace$ we find
\begin{align}
\frac{dU_\alpha}{du} &= \Big(g_{\rho\alpha}T^{\rho\mu\nu} - \left\lbrace \begin{smallmatrix} \mu\;\nu\\ \alpha \end{smallmatrix}\right\rbrace \Big)U_\mu U_\nu,
\end{align}
which is the final form of the geodesic equation. \\
\indent A particularly useful feature of geodesics in the case of a metric compatible connection is that the quantity $L = g^{\mu\nu}U_\mu U_\nu$ is constant along a geodesic. It turns out that this holds for our definition as well:
\begin{align*}
\frac{d}{du}\left(g^{\mu\nu}U_\mu U_\nu\right) &= \partial^\rho g^{\mu\nu} U_\rho U_\mu U_\nu + 2g^{\mu\nu}U_\nu \frac{d U_\mu}{du}\\
&= \left(\partial^\rho g^{\beta\nu}  + 2T^{\beta\rho\nu} - 2g^{\mu\beta}\left\lbrace \begin{smallmatrix} \rho\;\nu\\ \mu \end{smallmatrix}\right\rbrace \right)U_\beta U_\nu U_\rho\\
&= \left(\partial^\rho g^{\beta\nu}  - 2g^{\mu\beta}\left\lbrace \begin{smallmatrix} \rho\;\nu\\ \mu \end{smallmatrix}\right\rbrace\right)U_\beta U_\nu U_\rho\\
&= 0.
\end{align*}
This is extremely fortunate because it allows us to relate the distance function $D_{p(\tau)}^2(P,Q)$ directly to the integral $I(v)$, in particular
\begin{align}\label{synge}
I  = \frac{1}{2}D^2_{p(\tau)}(P,Q).
\end{align}

\section{Fourier Transform and its Inverse}\label{ap:technical}
In this appendix we explicitly verify some of the technical details discussed in the paper. Let us begin with equation \eqref{deltaP} which gives the Fourier representation for $\delta(p,q)$; the delta function on $\mc{P}$. Assuming $p,q \in \mc{D}_{p'}$ and $f(p) \in \mc{L}^2_{\hat{\mu}_{p'}}(D_{p'})$ we put
\begin{align}
\tilde{f}(q) &\equiv \int_{D_{p'}} d\mu_{p'}(p) \delta(p,q)f(p)\nonumber\\
&=\int_{D_{p'}} d\mu_{p'}(p) \int_{\mc{M}_{p'}} d\nu_{p'}(x) \mc{V}^{1/2}(p,p')\mc{V}^{1/2}(q,p')\nonumber\\
&\qquad \times\exp\left[ix^{\mu'}\left(\sig_{\mu'}^R(p,p') - \sig_{\mu'}^R(q,p')\right)\right]f(p).
\end{align}
The integral over $x$ covers the entire cotangent space $\mc{M}_{p'}$ and therefore turns the exponential into $\delta^{(4)}(\sig_{\mu'}^R(p,p') - \sig_{\mu'}^R(q,p'))$ which can be decomposed in the standard fashion. To do this we note that the uniqueness of the geodesic connecting $p'$ to $p$ and $p'$ to $q$ implies that $\sig_{\mu'}^R(p,p') =\sig_{\mu'}^R(q,p')$ if an only if $p = q$, and so
\begin{align}
 \delta^{(4)}\left(\sig^R_{\mu'}(p,p') - \sig^R_{\mu'}(q,p')\right)  =
\delta^{(4)}(p-q) \frac{ 
1}{|\mathrm{det}(\sig^{R \mu}{}_{\nu'})|} 
= \frac{\delta^{(4)}(p-q)}{\mc{V}(p,p') } \frac{\sqrt{g}_{p'} }{\sqrt{g}_{R_{p'}(p)}} ,
\end{align}
where the definition of the Van--Vleck Morette determinant together with  $ |{g}_{p'} \mathrm{det}(\sig^{R \mu}{}_{\nu'})| = |\mathrm{det}(\sig^{R \mu \nu'})|$ was used in the last equality. Substituting into our expression for $\tilde{f}(q)$ and noting that the presence of $\delta^{(4)}(p,q)$ allows us to replace all occurrences of $q$ with $p$ we find
\begin{align}
\tilde{f}(q) &= \int_{D_{p'}} d\mu_{p'}(p) \frac{\delta^{(4)}(p,q)}{\sqrt{g}_{R_{p'}(p)}}f(p)
=f(q),
\end{align}
where we used $p,q \in D_{p'}$ in the second equality. This demonstrates the validity of \eqref{deltaP} as a representation of the delta function. \\
\indent The Fourier representation of $\delta_{p'}(x,y)$, the ``delta function'' on $\mc{M}_{p'}$, is given in equation \eqref{deltaTP}. There are two important properties of this representation which we would like to verify: 
\begin{enumerate}
\item $\delta_{p'}(x,y)$ is a projector.
\item The image of $\delta_{p'}(x,y)$ is identical to the image of $\mc{F}_{p'}$.
\end{enumerate}
To demonstrate the first item recall \eqref{deltaTP},
\begin{align*}
\delta_{p'}(x,y)=\int _{D_{p'}}d\mu_{p'}(p)\mc{V}(p,p')\exp\left[i\sig_{\mu'}^R(p,p')\left(x^{\mu'}-y^{\mu'}\right)\right].
\end{align*}
 Making the change of variables $p_\mu \to Q'^{\mu} = \sig^{R\mu'}(p,p')$, the Jacobian of which is given in \eqref{changecor}, we find
\begin{align}
\delta_{p'}(x,y)=\int _{\Sigma_{p'}}\frac{d^4Q'}{\sqrt{g}_{p'}}\exp\left[iQ'_{\mu'}\left(x^{\mu'}-y^{\mu'}\right)\right],
\end{align}
where $D_{p'} \to \Sigma_{p'}$ under the coordinate change and $Q'_{\mu'}\equiv Q'^{\mu}g_{\mu'\nu'}(p')$. Unless $\Sigma_{p'} = \mathbb{R}^4$ the integral over $Q'_\mp$ does not give the usual  delta function $\delta^{(4)}(x-y)$. However, taking the convolution product of $\delta_{p'}$ with itself we find:
 \begin{align*}
 \int_{\mc{M}_{p'}} \rd\nu_{p'}(y) \delta_{p'}(x,y)\delta_{p'}(y,z) 
 &= \int_{D_{p'}\times D_{p'}} \frac{\rd^{4}Q' \rd^{4}K'}{|g_{p'}|} 
  \left(\int_{\mc{M}_{p'}} \rd\nu_{p'}(y) e^{i y^{\nu'}(Q'_{\mu'}-K_{\mu'})} \right)  \\
  &\qquad \times e^{i K_{\mu'}x^{\mu'}}e^{-iQ'_{\mu'}z^{\mu'}}\nonumber\\ 
  &=  \int_{D_{p'}\times D_{p'}} \frac{\rd^{4}Q' \rd^{4}K'}{\sqrt{g}_{p'}} 
 {\delta^{(4)}(Q'-K')}
  e^{i K_{\mu'}x^{\mu'}} e^{-iQ'_{\mu'}z^{\mu'}}\nonumber\\ 
  &= \delta_{p'}(x,z),
 \end{align*}
 which confirms that $\delta_{p'}(x,y)$ is a projector, i.e. identity onto its image. For the second item, suppose $\hat{f}_{p'}(x) \in  \mc{F}_{p'}(\mc{L}^2_{\hat{\mu}_{p'}}(D_{p'}))$ so there exists a function $f(p) \in \mc{L}^2_{\hat{\mu}_{p'}}(D_{p'})$ such that
\begin{align}
\hat{f}_{p'}(x) = \int_{D_{p'}} d\mu_{p'}(p) \mathcal{V}^{1/2}(p,p')\exp\left(-ix^{\mu'}\sig^R_{\mu'}(p,p')\right)f(p).
\end{align}
Evaluating the convolution of $\delta_{p'}$ with $\hat{f}_{p'}$ we find
\begin{align*}
(\delta_{p'}\circ \hat{f}_{p'})(x) &=\int_{\mc{M}_{p'}} d\nu_{p'}(x)\int _{D_{p'}}d\mu_{p'}(p)\mc{V}(p,p')\exp\left[i\sig_{\mu'}^R(p,p')\left(x^{\mu'}-y^{\mu'}\right)\right]\\
&\qquad\times \int_{D_{p'}} d\mu_{p'}(q) \mathcal{V}^{1/2}(q,p')\exp\left(-ix^{\mu'}\sig^R_{\mu'}(q,p')\right)f(q)\\
&=\int _{D_{p'}}d\mu_{p'}(p)d\mu_{p'}(q)\mc{V}^{1/2}(p,p')\exp\left(-i\sig_{\mu'}^R(p,p')y^{\mu'}\right)\\
&\qquad\times \int_{\mc{M}_{p'}} d\nu_{p'}(x)\mc{V}^{1/2}(p,p') \mathcal{V}^{1/2}(q,p')\exp\left[ix^{\mu'}\left(\sig^R_{\mu'}(p,p') - \sig^R_{\mu'}(q,p')\right)\right]f(q)\\
&=\int _{D_{p'}}d\mu_{p'}(p)d\mu_{p'}(q)\mc{V}^{1/2}(p,p')\exp\left(-i\sig_{\mu'}^R(p,p')y^{\mu'}\right)\delta(p,q)f(q)\\
&=\int _{D_{p'}}d\mu_{p'}(p)\mc{V}^{1/2}(p,p')\exp\left(-i\sig_{\mu'}^R(p,p')y^{\mu'}\right)f(p)\\
&= \hat{f}_{p'}(x),
\end{align*}
where we have used the Fourier representation of $\delta(p,q)$ in going from the third line to the fourth. This shows that the image of $\delta_{p'}$ under convolution is identical with the image of $\mc{F}_{p'}$, verifying the second item above.

\bibliography{QFT_Final}{}
\bibliographystyle{unsrt}
\end{document}